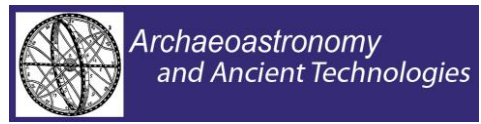

# Ancient astronomical instrument from Srubna burial of kurgan field Tavriya-1 (Northern Black Sea Coast)

## Larisa Vodolazhskaya[1], Pavel Larenok[2], Mikhail  Nevsky[3]

[1] Southern Federal University (SFU), Rostov-on-Don, Russian Federation;
E-mails: larisavodol@aaatec.org, larisavodol@yahoo.com
[2] NP "Yuzharheologiya", Rostov-on-Don, Russian Federation; E-mail: dao2@inbox.ru
[3]Southern Federal University (SFU), Rostov-on-Don, Russian Federation; E-mails:
munevsky@sfedu.ru

### Abstract

The article presents the results of analysis of the spatial arrangement of the wells on the unique slab from Srubna burial of kurgan field Tavriya-1 (Rostov region, Russia) by astronomical methods. At the slab revealed two interrelated groups of wells, one of which - in the form of a circle, is proposed to interpret how analemmatic sundial, and second group, consisting of disparate wells, as auxiliary astronomical markers of rising luminaries directions, to correct the position of the gnomon. Simultaneous location of both groups of wells on the same slab is a possible indication of one of the stages of development of the design features analemmatic sundial - setting movable gnomon and technology of measuring time with it. It may point to local origin, as the very idea of analemmatic sundial as well technology measurement of time with them.

The article also describes the model analemmatic sundial, hour marks which in many cases coincide with the wells arranged in a circle, particularly in a working range from 6 to 18 hours. In the study proposed a method which can identify moments of solstices and equinoxes in ancient times with the help of the gnomon of analemmatic sundial and mobile gnomons, installed in wells belonging to the second group. The opportunity of use analemmatic sundial as moondial in a full moon night. Slab with two groups of wells is proposed to consider, as the oldest astronomical instrument discovered in the Northern Black Sea coast, which allowed to observe the apparent motion of the Sun and the Moon and allowed measure the time during the day, using analemmatic sundial and at night during the full Moon - with the help of moondial.

## Introduction

In 1991-1992 the Taganrog archaeological expedition under the leadership of P.A. Larenok were conducted archaeological research of kurgan field Tavriya - 1 of Neklinovsky district of Rostov region near the farm Tavrichesky [1]. During excavations in the Srubna burial 2 of kurgan 1 of kurgan field Tavriya - 1 was found unique slab, which were knocked out two circles of wells (Figure 1). Thanks to these circles of wells the slab was been classified as altar. Then E.I. Bespaly and P.A. Larenok were first suggested to the astronomical appointment of wells circles.

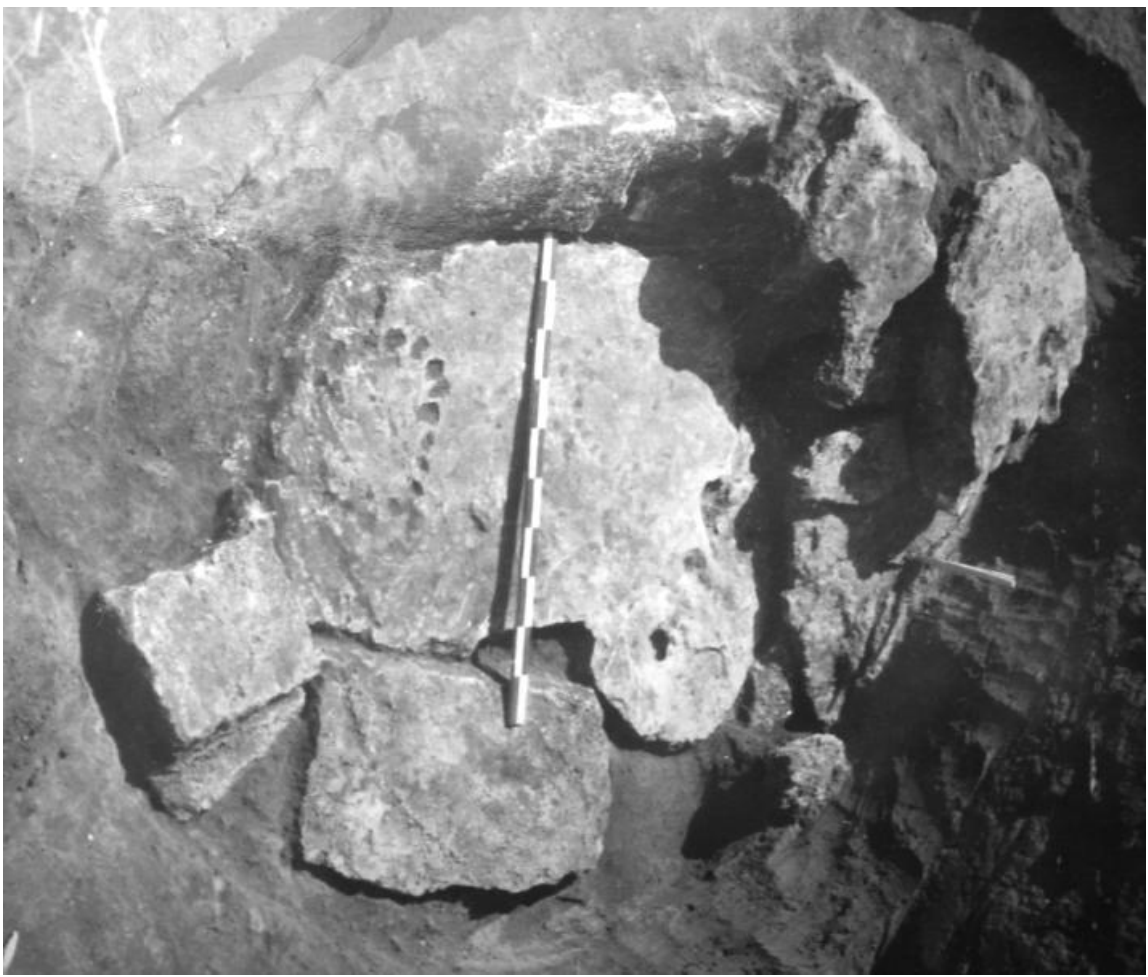

**Figure 1**. Kurgan field Tavriya - 1, kurgan 1, burial 2, slab with the wells on-site detection. View from the North (photo by Larenok P.A., 1991).

Kurgan 1 was a round in plan, slightly elongated in the meridional direction, up the hill with a diameter of about 20-30 m and a height of approximately 2.2 m. Kurgan was built on a natural elevation above two burials of the Bronze Age, in the overlap of one of them - the burial 2 - slab with wells was found. Depth of burial 2 was 3.55 m. Burial pit was closed of complex cluster of stones, which was traced with a depth of 2 - 2.15 m. Cluster was wrong oval: 2.5 x 1.8 m. It was the long axis oriented NE - SW. Cluster consisted of a central slab, overlapping grave pit. Its dimensions are approximately: 1.45 x 1.0 x 0.15 m. Along the perimeter of the burial pit was made ring fence, consisting of small platy stones stacked in 2-3 rows vertically. Stones of fence partially slipped on the central plate. North and south from the cluster of stones at the level of the buried soil traced clay mainland dozer blade. Depth - 2.30 - 2.55 m. Stones of fence partially overlapping dozer blade. The central cluster plate, on the outer side thereof, are fixed two circles composed of round wells. The diameter of the circle was about 0.3 - 0.4 m. Diameter wells - 3-4 cm.

Grave pit, except plate, overlaps another cluster of reeds and wood, ashes which was marked under the upper stone of fences. At the bottom of the burial pit traced the remains of reed mats.

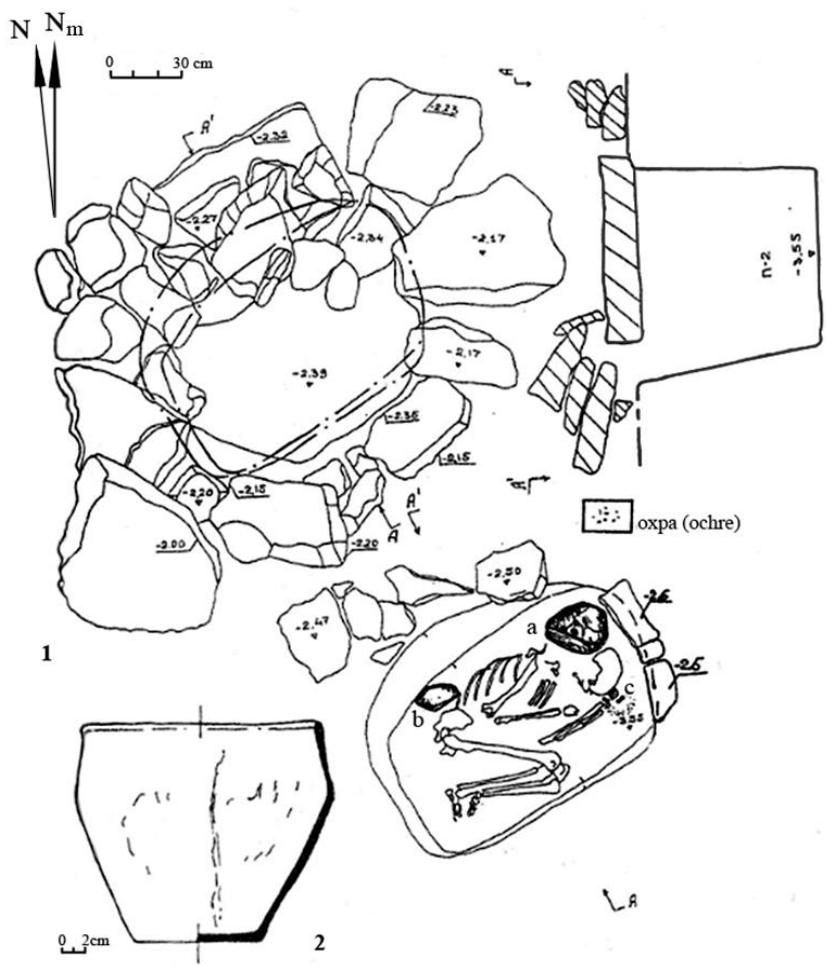

**Figure 2**. Kurgan field Tavriya - 1, kurgan 1: *1* – burial 2, *a* - vessel, *b* - stone - limestone of irregular shape, *c* – astragal of ram; *2* - burial 2, vessel. The dotted line indicates the approximate outline of the floor slab with the wells. N - True North[3], Nm - Magnetic North.

The backbone of the men about 50 years old lay crouched on the left side, the skull is oriented NE. Bones of the arms bent at the elbows, his left hand - in front of the front part of

---

the skull, right hand - on the chest in front of the skull. Bones of the legs are bent to the left. Per skull with NW lay vessel (Fig. 2a). Surface is grayish-brown, dark and sooty. Potsherd in a kink is black, shallow fireclay and limestone inclusions are in the test. On edge of the vessel - an ornament in the form of a horizontal row of finger impressions. Ashes of burnt organic mass traced in a vessel. Astragals of ram were found in the left hand from the south (see Fig. 2c). One more astragal was in filling the grave pit. Over the bones of the pelvis to the NW was a stone - limestone of irregular shape (Fig. 2b). Ochre stains were noted at the bottom of the burial pit: under vessel, under the brush of his left hand. Just brown spots of decay were observed under the bones.

Slab of light gray sandstone is irregular weakly expressed V - shaped, with chipped, roughly rounded corners. The largest slab size 85 x 164 cm, thickness about 20 cm. Wells located on one side of the plate. Initially, the surface of the slab were knocked two circles from the wells. However, because of the storage slab out of premises, a large part of one of the circles, unfortunately, has been lost, and the plate got a crack. Plate was in a protected area of the Archaeological Museum "Tanais", so it is protected from minor damages that could inflict ordinary visitors. The occurrence of cracks was associated with random collisions on the plate of the car, and a large circle was damaged (scraped off) in an unsuccessful maneuvering the tractor close to the plate. Now we can see large circle of wells only on the photo 1991 (see Fig. 1). Spite of the damages, many of the wells on the slab are well enough distinguish and amenable to analysis (see Fig. 3).

The aim of our study was to analyze the spatial arrangement of the wells on the slab of Srubna burial 2 of kurgan 1 of kurgan field Tavriya – 1 by astronomical methods.

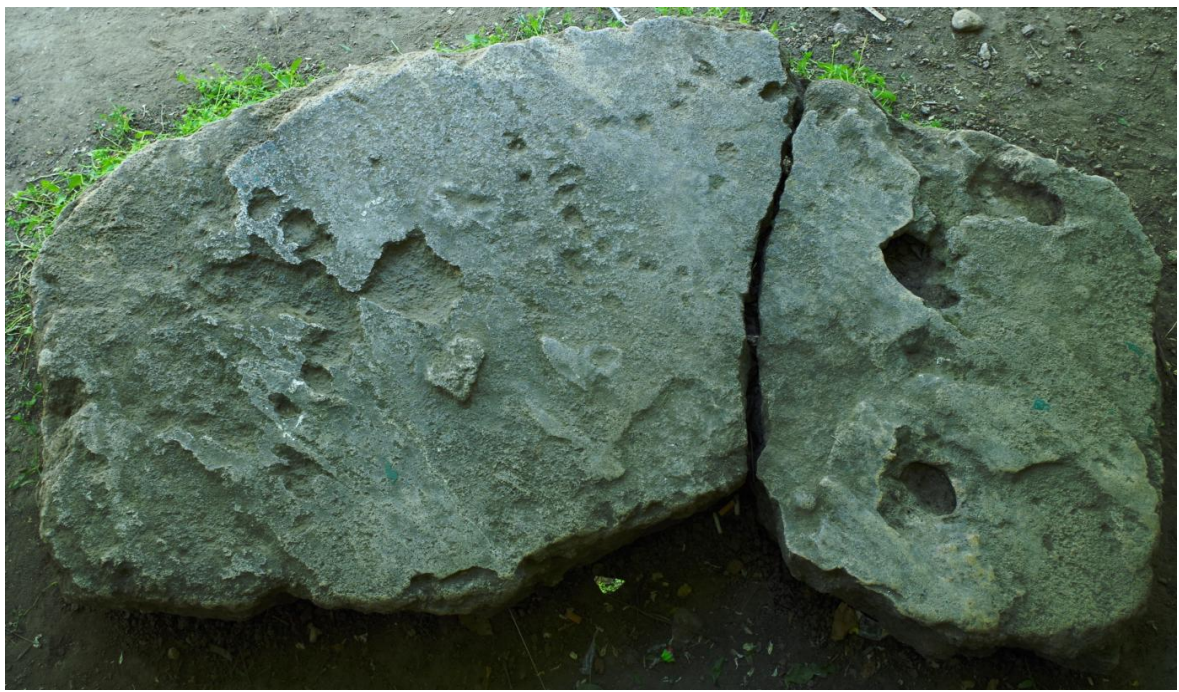

**Figure 3**. Kurgan field Tavriya - 1, kurgan 1, burial 2, slab with the wells (photo by Vodolazhsky D.I., 2014).

The relevance of the study is related to the uniqueness of the detected plate having a complex set of the wells on its surface.

As a result of our study was identified relationship surviving circular group the wells and groups wells arranged linearly, as well as their relationship with astronomically significant directions. We have proposed a model analemmatic sundial describing wells arranged in a circle, as well as the technology of determining moments of the solstices and equinoxes using gnomon of analemmatic sundial and additional moving gnomons installed in wells, belonging to the second group.

**Analemmatic sundial**

Cupped signs – wells, inflicted, including, in a circle, find on all continents. However, so far there is no consensus about their interpretation. Cupped signs are often found on slabs of cysts, passage graves, dolmens and ancient cemeteries. The wells are located in a circle, reminiscent of marks of sundial, and in European countries, there is still an ancient tradition to place sundial on the graves [2].

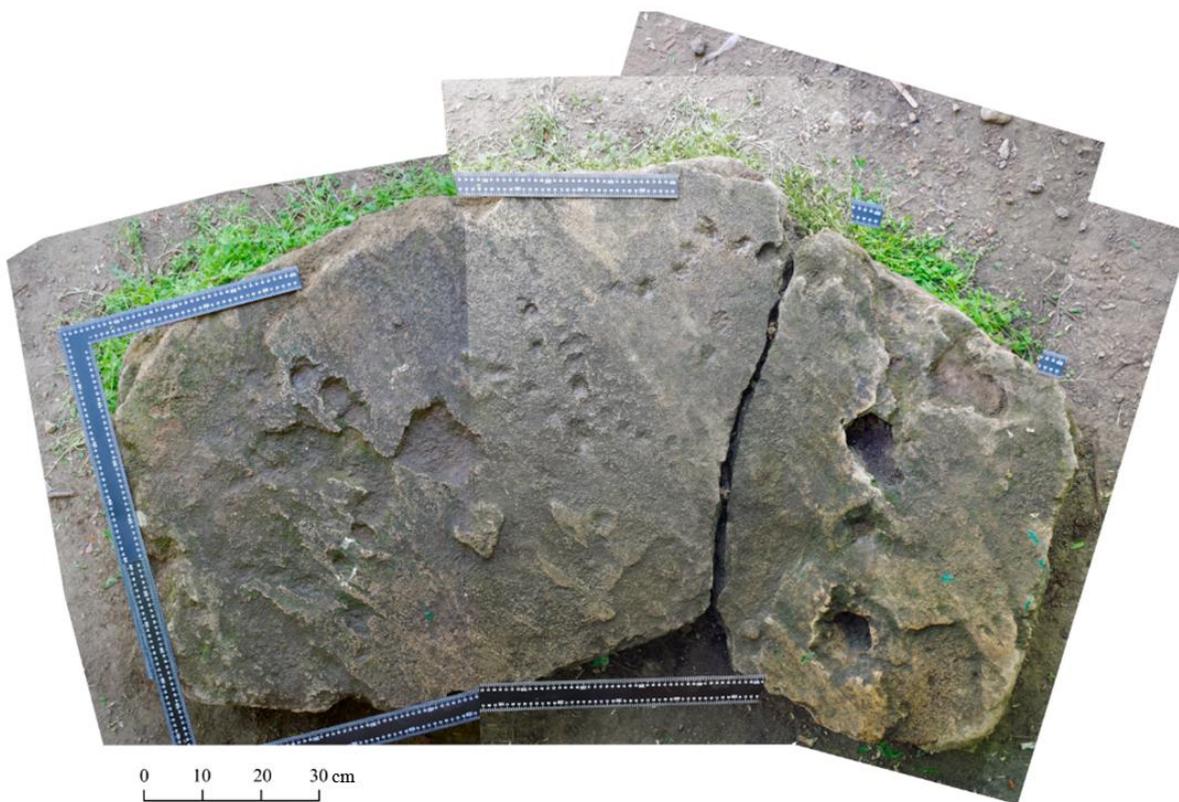

*a*

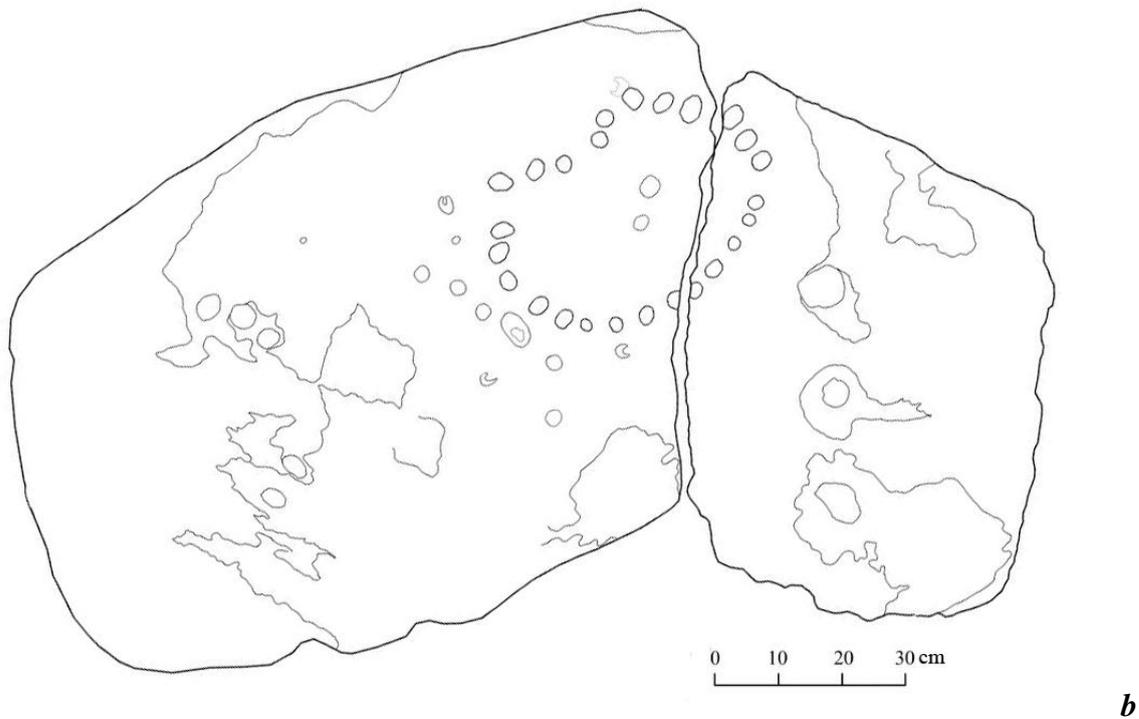

**Figure 4**. Kurgan field Tavriya - 1, kurgan 1, burial 2, slab with the wells: *a* - the result of "gluing" photo-fragments; *b* – drawing of the surface of the slab.

It is known that the image of a sundial was found in the tomb of Seti I (around 1300 BC) in Egypt [3]. In the Valley of the Kings in 2013 have been found vertical sundial dating from the XIII century BC [4], which have been marked in accordance with the division of the day at the 24 hour equal duration [5], [6].

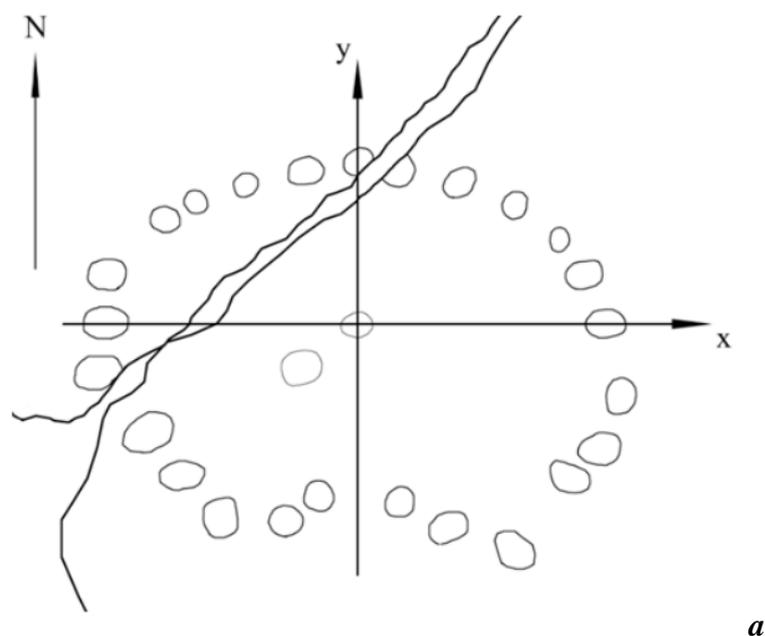

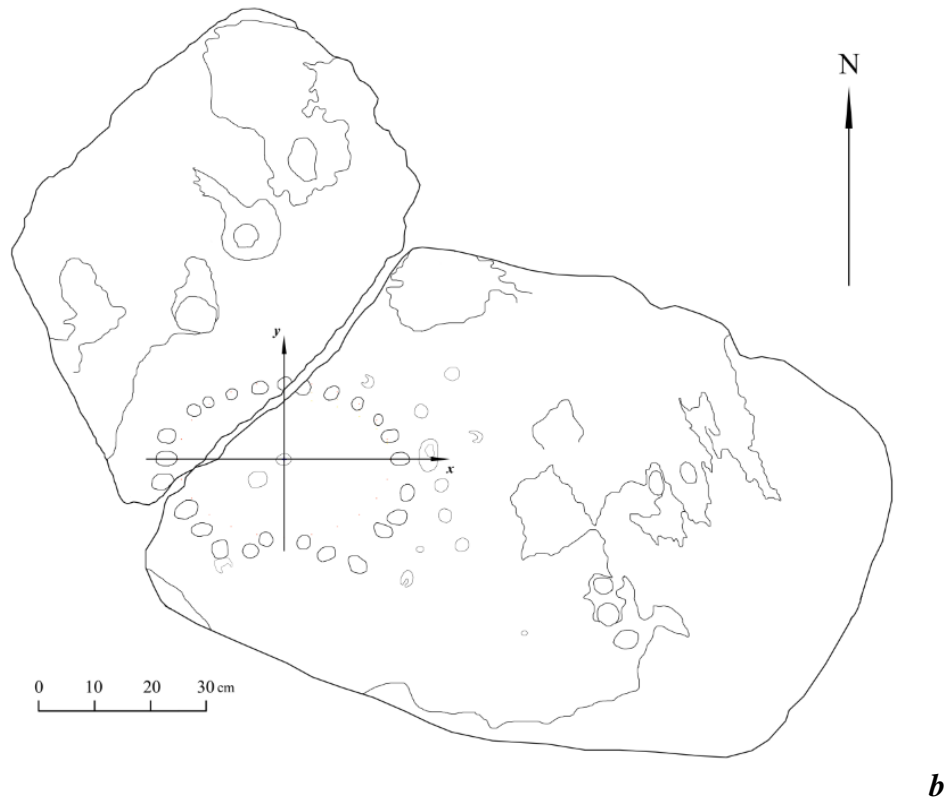

***b***

**Figure 5**. Kurgan field Tavriya - 1, kurgan 1, burial 2, slab with the wells: ***a*** - ellipse of the wells with X and Y axes for analemmatic sundial; ***b*** – slab with wells oriented according to sides of the world in line with the axes X and Y. N - True North.

Plate from Tavriya-1dates from the XV-XIII century BC - the time close dating sundial in Egypt. Therefore, we can assume that into said era could exist transfer ("migration") of technology of sundial markup among the peoples, inhabiting coast of the Mediterranean and Black Seas. For the XIII century BC were characterized by processes of migration of the peoples of the Balkans and Asia Minor up to Egypt, have received the name Sea Peoples migration [7]. However, the preponderance of evidence of participation of the population of the Northern Black Sea coast in these migration processes still were not found. As positive evidence can be considered mention of participation of the Amazons, placed by Herodotus in the Northern Black Sea coast in the "three days' journey from Tanais" (Herodotus. History IV, 110 - 116), in the Trojan War (Scholiast on Homer, Iliad. XXIV. 804; Apollodorus , Epitome 5.1; Proclus, Chrestomathy 2, Aethiopis), participation in which taking the Sea Peoples [8].

Wells on the slab from Tavriya-1 were arranged in a circle, so we assumed that they could be the hour markers of analemmatic sundial similarly slab with wells from Popov Yar [9].

For the proof that the find is indeed a sundial, we have conducted interdisciplinary research with the help of natural science methods. In recent decades, the complex interdisciplinary studies are widely used in history and archeology. They led to the emergence of new scientific fields such as archaeoastronomy, historical informatics, etc. [10 - 15].

With the help of the camera *Pentax K-50*, we performed photographing surface of the slab with respect to squares 40 x 60 cm with a height of one and half meters at an angle of lens $90^0$ to the surface. Selected parameters of photography helped to minimize the distortions that do not exceed 0.5 cm for every 60 cm. Next, in the graphical editor, we performed "gluing" photo-fragments into one seamless image (see Fig. 4).

Circle of the wells on the slab from Tavriya-1 has two protrusions - "cat ears". We assumed that they were made to determine the axis of symmetry. In accordance with this assumption, we conducted a vertical axis Y between these two protrusions. Thus, the axis passed through a central the wells of circle. Having taken it for a possible center of analemmatic sundial, which was mounted gnomon at the equinoxes, we conducted there through horizontal axis X. Since the working part of analemmatic "the dial" must be an ellipse, we related the "cat ears" with its southern part - not a working part (see Fig. 5).

To the upper part of the ellipse the measured semi-major axis of the ellipse (East), is $M_1 \approx 19$ cm on the inner edge of the well and $M_2 \approx 22$ cm on the outer edge of the well. Measurement of semi-minor axis interfered crack. Measured semi-minor axis (North) is $m_1 \approx 12.5$ cm on the inner edge of the well and $m_2 \approx 14.8$ cm on the outer edge of the well. In analemmatic sundial these quantities are related. Knowing the semi-major axis of the ellipse *M,* we can calculate by the formula 1 the semi-minor axis m. For the semi-major axis $M = 19.0$ cm calculated semi-minor axis is $m = 14.0$ cm. This value is approximately equal to the measured value, that the testifies in favor of that wells on the slab from Tavriya-1 could be the hour markers of analemmatic sundial.

Coordinates of the hour marks and the coordinates of the gnomon for analemmatic sundial, marked in accordance with the division day to 24 hours of equal duration, are calculated as follows [16]:

$$m = M \cdot \sin \varphi \quad (1)$$

$$x = M \cdot \sin H \quad (2)$$

$$y = M \cdot \sin \varphi \cdot \cos H \quad (3)$$

$$Z_{ws} = M \cdot tg\, \delta_{ws} \cdot \cos \varphi \quad (4)$$

$$Z_{ss} = M \cdot tg\, \delta_{ss} \cdot \cos \varphi \quad (5)$$

$$H^{/} = arctg\left( \frac{tgH}{\sin \varphi} \right) \quad (6)$$

$$H = 15^0 \cdot (t - 12)$$

where *x* - coordinate of a point on the *X* axis for analemmatic sundial, *y* - coordinate of the point on the *Y* axis for analemmatic sundial, $M \approx 19$ cm - measured semi-major axis of the ellipse, $\varphi$ - latitude of location, *H* - hour angle of the Sun, $H^{/}$ - angle between the meridian line and the hour line on the sundial, $\delta_{ws} = -\varepsilon$ - declination of the Sun at the winter solstice, $\delta_{ss} = \varepsilon$ - declination of the Sun at the summer solstice, $y = Z_{ws}$ – in the winter solstice, $y = Z_{ss}$ – in the summer solstice (Fig. 6 ).

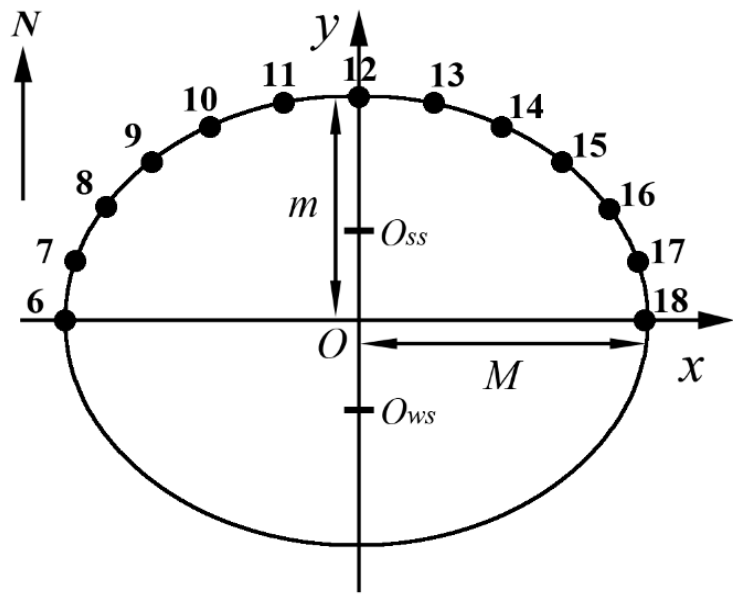

**Figure 6**. The coordinate plane with the hour markers from 6 to 18 hours. *M* - semi-major axis of the ellipse, m - minor semi axis of the ellipse, *O* - center of the ellipse, $O_{ws}$ - the position of the gnomon at the winter solstice on analemmatic sundial, *Oss* - the position of the gnomon at the summer solstice for analemmatic sundial. N - True North.

The results of our calculations of the coordinates x and y of the hour marks of analemmatic sundial by formulas 2 and 3 for geographic coordinates of Tavriya-1 *Lat*=47°17′N *Long*=38°37′E are given in Table 1. Calculated according to the formula 4 of gnomon position at the winter solstice $Z_{ws}$=-5.7 cm calculated by the formula 5 of gnomon position at the summer solstice $Z_{ss}$=5.7 cm. The gnomon is at the center of coordinates at the equinox.

For comparison the location of the wells with analemmatic sundial with semi-major axis of the same length *M*≈19 cm, but for other latitudes, we have calculated the hour markers for latitudes Lat = 41°00′ N (tab. 2) and Lat = 35°00′ N (tab. 3). The location of all the calculated hour markers relatively the wells slabs is shown in Figure 7.

**Table 1.** Coordinates of the hour marks of analemmatic sundial for *Lat* = 47°17′ N. *H* − the hour angle of the Sun, *H′*- angle between the meridian line and the hour line on the sundial, *t* - time, *x* - coordinate of the mark on the axis *X*, *y* - coordinate of the mark on the axis *Y*.

| | t, (hour) | | | | | | | | | | | | |
|---|---|---|---|---|---|---|---|---|---|---|---|---|---|
| | 6 | 7 | 8 | 9 | 10 | 11 | 12 | 13 | 14 | 15 | 16 | 17 | 18 |
| H, (0) | -90.0 | -75.0 | -60.0 | -45.0 | -30.0 | -15.0 | 0.0 | 15.0 | 30.0 | 45.0 | 60.0 | 75.0 | 90.0 |
| H', (0) | -90.0 | -78.9 | -67.0 | -53.7 | -38.2 | -20.0 | 0.0 | 20.0 | 38.2 | 53.7 | 67.0 | 78.9 | 90.0 |
| x, (cm) | -19.0 | -18.4 | -16.5 | -13.4 | -9.5 | -4.9 | 0.0 | 4.9 | 9.5 | 13.4 | 16.5 | 18.4 | 19.0 |
| y, (cm) | 0.0 | 3.6 | 7.0 | 9.9 | 12.1 | 13.5 | 14.0 | 13.5 | 12.1 | 9.9 | 7.0 | 3.6 | 0.0 |

**Table 2.** Coordinates of the hour marks of analemmatic sundial for *Lat* = 41°00′ N. *H* − the hour angle of the Sun, *H′*- angle between the meridian line and the hour line on the sundial, *t* - time, *x* - coordinate of the mark on the axis *X*, *y* - coordinate of the mark on the axis *Y*.

| | t, (hour) | | | | | | | | | | | | |
|---|---|---|---|---|---|---|---|---|---|---|---|---|---|
| | 6 | 7 | 8 | 9 | 10 | 11 | 12 | 13 | 14 | 15 | 16 | 17 | 18 |
| H, (0) | -90.0 | -75.0 | -60.0 | -45.0 | -30.0 | -15.0 | 0.0 | 15.0 | 30.0 | 45.0 | 60.0 | 75.0 | 90.0 |
| H', (0) | -90.0 | -80.0 | -69.3 | -56.7 | -41.3 | -22.2 | 0.0 | 22.2 | 41.3 | 56.7 | 69.3 | 80.0 | 90.0 |
| x, (cm) | -19.0 | -18.4 | -16.5 | -13.4 | -9.5 | -4.9 | 0.0 | 4.9 | 9.5 | 13.4 | 16.5 | 18.4 | 19.0 |
| y, (cm) | 0.0 | 3.2 | 6.2 | 8.8 | 10.8 | 12.0 | 12.5 | 12.0 | 10.8 | 8.8 | 6.2 | 3.2 | 0.0 |

**Table 3.** Coordinates of the hour marks of analemmatic sundial for *Lat* = 35°00′ N. *H* − the hour angle of the Sun, *H'*- angle between the meridian line and the hour line on the sundial, *t* - time, *x* - coordinate of the mark on the axis *X*, *y* - coordinate of the mark on the axis *Y*.

| | t, (hour) | | | | | | | | | | | | |
|---|---|---|---|---|---|---|---|---|---|---|---|---|---|
| | 6 | 7 | 8 | 9 | 10 | 11 | 12 | 13 | 14 | 15 | 16 | 17 | 18 |
| H, (0) | -90.0 | -75.0 | -60.0 | -45.0 | -30.0 | -15.0 | 0.0 | 15.0 | 30.0 | 45.0 | 60.0 | 75.0 | 90.0 |
| H', (0) | -90.0 | -81.3 | -71.7 | -60.2 | -45.2 | -25.0 | 0.0 | 25.0 | 45.2 | 60.2 | 71.7 | 81.3 | 90.0 |
| x, (cm) | -19.0 | -18.4 | -16.5 | -13.4 | -9.5 | -4.9 | 0.0 | 4.9 | 9.5 | 13.4 | 16.5 | 18.4 | 19.0 |
| y, (cm) | 0.0 | 2.8 | 5.4 | 7.7 | 9.4 | 10.5 | 10.9 | 10.5 | 9.4 | 7.7 | 5.4 | 2.8 | 0.0 |

Hour lines of analemmatic sundial, according to calculated angles and markers (tab. 1), were applied to the drawing of fragment of slab with wells (Fig. 7). Center of the hour lines is the site of the gnomon attachment in the equinox and corresponds to the center of the ellipse - the point *O* with coordinates *(0;0)*. Analemmatic sundial gnomon is a vertical rod which is moved along the *Y* axis, between the points $O_{ws}$ and $O_{ss}$. This gnomon was easily interchanged as a chess piece. Hour lines for equinox highlighted in red in Figure 7. Coordinates of the ends of these lines − red points - have coordinates calculated hour markers for latitude Tavriya-1 (Tab. 1). The dotted lines in Figure 7 are hypothetical hour lines in the range from 18 to 6 am, because this time range is not working in the equinox. In figure 7 orange dots noted the location of hour markers for latitude *Lat* = 41°00' *N* and yellow dots − for latitude *Lat* = 35°00' *N*.

In Figure 7 shows that the hour marks corresponding to the latitude of the detection slab *Lat* = 47°17' *N* good coincide (primarily in the working range from 6 to 18 hours on true solar time[4]) with wells on the slab and coincide much better than marks calculated for the more southerly latitudes. This coincidence also confirms the assumption that wells on the slab from Tavriya-1 are hour markers of analemmatic sundial made at the latitude of detection slab.

---

[4] Local true solar time does not coincide with the official time of a country or an area.

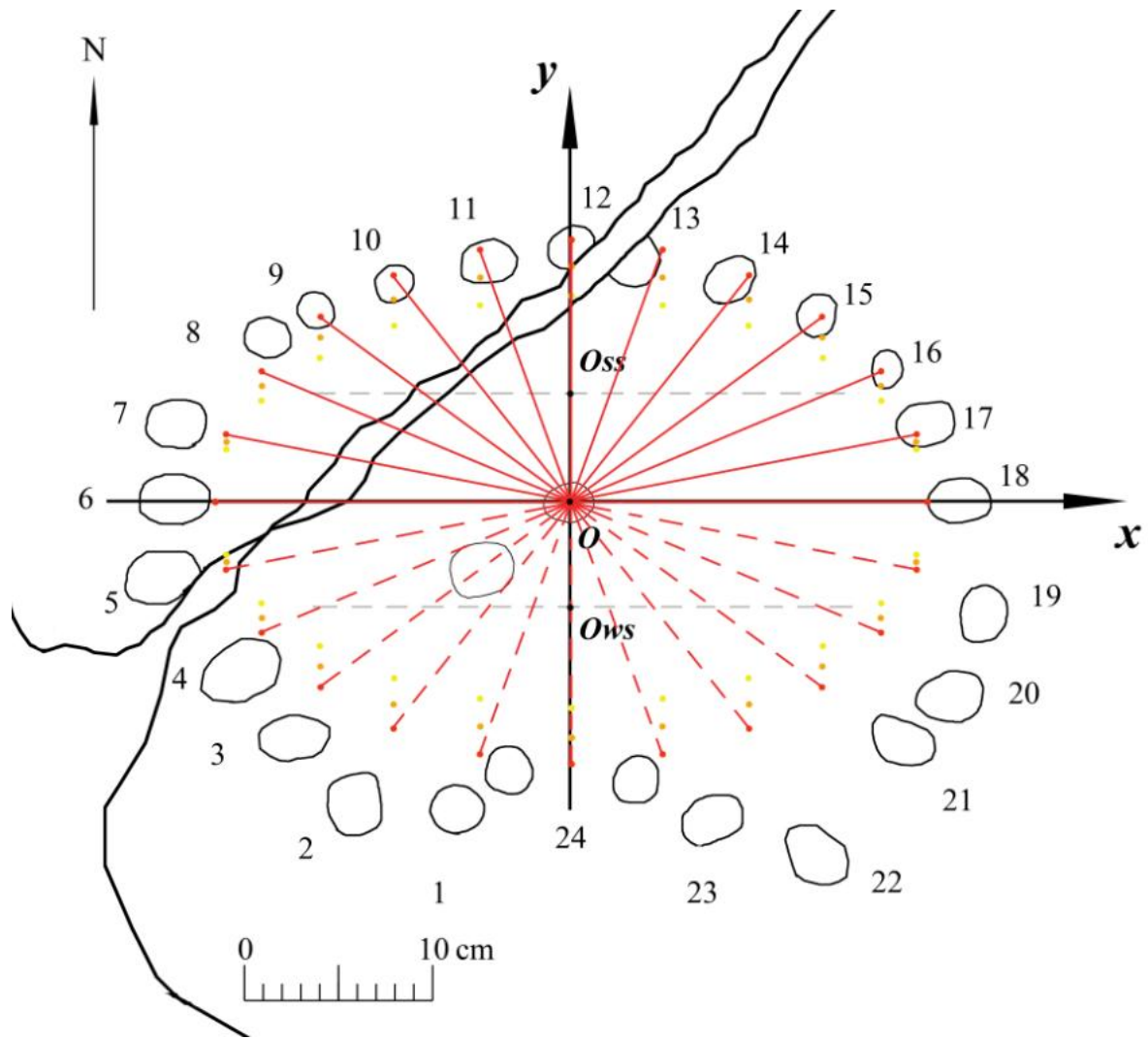

**Figure 7**. Kurgan field Tavriya - 1, kurgan 1, burial 2, slab with the wells: fragment with wells with applied hour lines of analemmatic sundial of latitude Tavriya-1 for equinox (red color). Hour markers for latitude 41°00' *N* are marked in orange color, and for the latitude 35°00' *N* – yellow color. *N* - True North .

For proper operation of analemmatic sundial for the latitude 47°17' *N* at the summer solstice the gnomon of analemmatic sundial must be installed on the line *SS*, passing through the point *Oss* (0; 5.7), and at the winter solstice - on line *WS*, passing through point *Ows* (0; -5.7) (see Fig. 6). Daylength in the summer is the biggest, so hour lines include a maximum time range - from 4 to 20 hours. Daylength in winter shorter, therefore hour lines will cover a smaller time range - from about 8 to 16 hours (see Fig. 8).

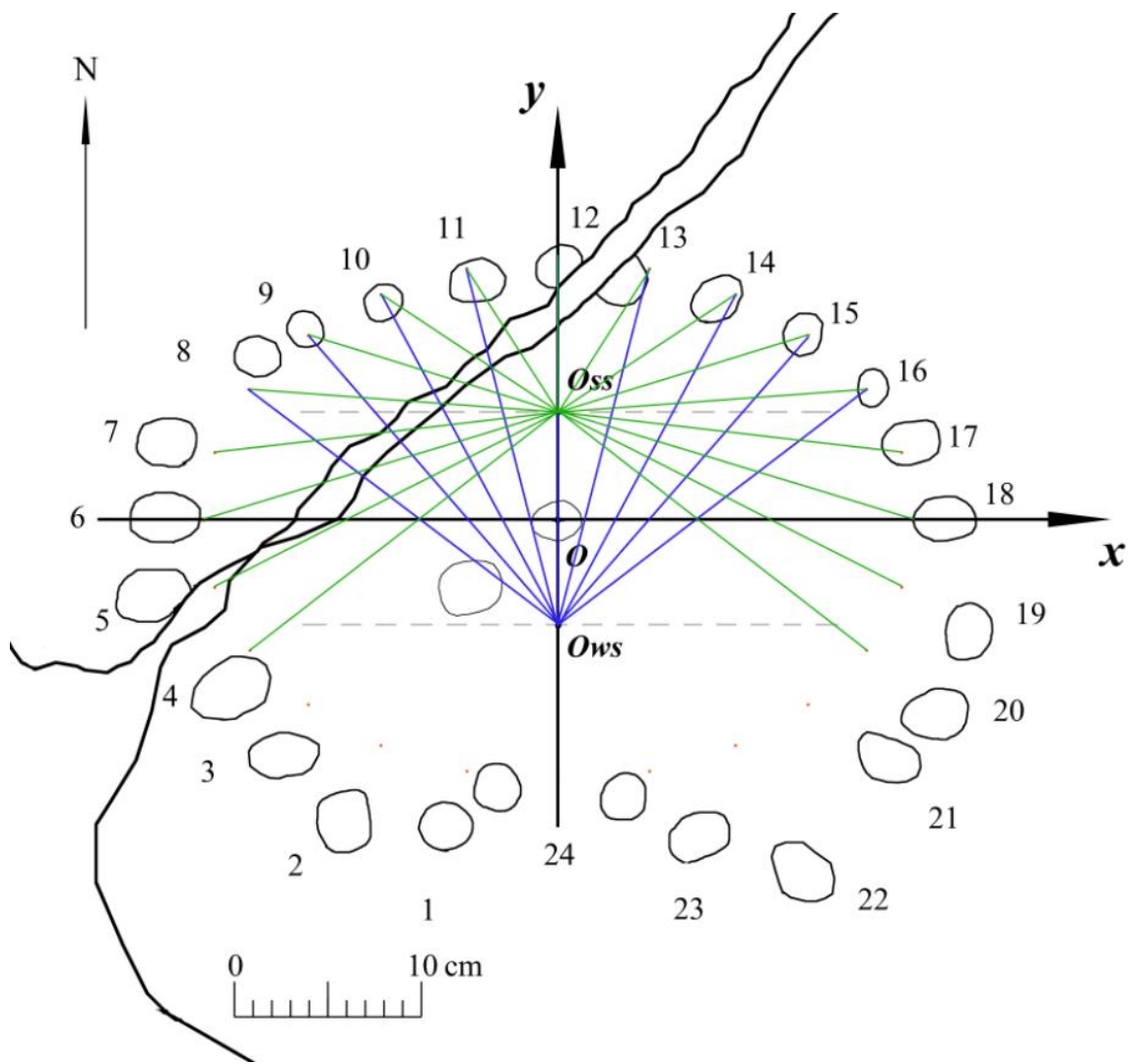

**Figure 8**. Kurgan field Tavriya - 1, kurgan 1, burial 2, slab with the wells: fragment with wells with applied hour lines of analemmatic sundial of latitude Tavriya-1; blue line - hour line at the winter solstice, the green line - hour line at the summer solstice. N - True North.

In the case of a slab from Tavriya-1 line, passing through the point *Ows*, almost touches the second round well located inside the ellipse. Thus, this well (southern edge) serves as a marker for the line, passing through the point *Ows*, in which must be installed gnomon of analemmatic sundial in the winter solstice. Such a coincidence is also evidence in favor of the interpretation of the circular wells on the plate from Tavriya-1 as analemmatic sundial.

**Astronomical markers**

To the east of the circle of wells on the surface of the slab can be discerned still few scattered wells, most of which located linear. We assumed that these wells were connected with astronomically important directions and performed the auxiliary function with respect to analemmatic sundial.

The point of intersection of the Moon orbit with the Ecliptic (in the projection of the celestial sphere) - the nodes of the lunar orbit, constantly shifted along the Ecliptic, describing a full circle in 18.6 years. Declination of the Moon changes during the sidereal month[5] between the two extremes values. When the ascending node of the lunar orbit (where the Moon passes from the southern part of the sky to the north) is the same with the point of the vernal equinox, the declination of the Moon can reach $\pm\delta=\varepsilon+i$ (High Moon). It happens once in 18.6 years. In the middle of this interval, when the descending node of the lunar orbit (where the Moon passes from the northern to the southern sky) coincides with the point of the vernal equinox, the declination of the Moon can reach $\pm\delta=\varepsilon-i$ (Low Moon). For example, High or Low Moon on a full moon in the winter in moment of culmination can occupy more or less high position above the horizon, respectively [17], [18].

Declination of High Moon at the full moon near the summer solstice take $\delta=\varepsilon+i$, near the winter solstice $\delta=-\varepsilon-i$, near the equinox $\delta=i$. Declination of Low Moon at the full moon near the summer solstice take $\delta=\varepsilon-i$, near the winter solstice $\delta=-\varepsilon+i$, near the equinox $\delta=-i$.

Azimuth calculation of sunrise and sunset were made by the formulas [19]:

$$\cos A_r = \frac{\sin\delta - \sin\varphi \cdot \sin h}{\cos\varphi \cdot \cos h}, \qquad (7)$$

$$A_s = 360^0 - A_r, \qquad (8)$$

where $A_r$ - azimuth of rise, measured from north to east (surveyor), $A_s$ - azimuth of set, $\delta$ - declination, $h$ - altitude, $\varphi$ - latitude. Calculations are made on the upper edge of the disk:

$$h = -R - \rho + p \quad, \qquad (9)$$

where $R$ - 1/2 angular size, $\rho$ - refraction at the horizon, $p$ - horizontal parallax [20]. For the Sun and the Moon $R=16'$, $\rho=35'$ [21]. For the Sun $l=1.496 \times 10^{11}$ м, $p=8.8''$. For the Moon $l=3.844 \times 10^8$ m, $p=57'$ [22].

During summer solstice the Sun declination equal to angle of ecliptic inclination to celestial equator $\varepsilon$, which is calculated using the formula [23]:

$$\varepsilon = 23.43929111^0 - 46.8150'' \cdot T - 0.00059'' \cdot T^2 + 0.001813 \cdot T^3 \qquad (10)$$

$$T \approx \frac{(y - 2000)}{100} \qquad (11)$$

where $T$ - the number of Julian centuries, that separates this age from noon of the 1 of January 2000, $y$ - year of required age. During winter solstice the Sun declination $\delta = -\varepsilon$, and during equinoxes $\delta = 0$. The Moon orbit plane is inclined to ecliptic at angle $i \approx 5.145^0$.

Calculated by us according to the formula 10 tilt angle of the ecliptic to the celestial equator $\varepsilon=23^0 50' 20''$ for 1200 BC. The results of our calculations of the azimuth of sunrise and sunset according to the formulas 7 - 9 for astronomically important events are shown in Table 4.

---

[5] The time interval between two successive returns of the Moon in the same (relative to the stars) the place of the celestial sphere.

**Table 4.** Azimuths of the Sun at sunrise (sunset) at the equinoxes and solstices, calculated on the top edge of the visible disk of the Sun; *h* - height of the Sun, *δ* - declination of the Sun, *A* - the azimuth of the Sun.

| phenomenon | h, (0) | δ, (0) | A, (0) | The letter designation 6 |
|---|---|---|---|---|
| summer solstice, sunrise | -0.85 | 23.84 | 52.27 | - |
| equinox, sunrise | -0.85 | 0.00 | 89.08 | A1 |
| winter solstice, sunrise | -0.85 | -23.84 | 125.44 | A21, A22 |
| summer solstice, sunset | -0.85 | 23.84 | 307.73 | - |
| equinox, sunset | -0.85 | 0.00 | 270.92 | - |
| winter solstice, sunset | -0.85 | -23.84 | 234.56 | A3 |

The results of calculations of azimuths of rise and set of High and Low Moon according to the formulas 7 - 9 are shown in Table 5.

**Table 5.** Azimuths of High and Low Moon at the time of Moon rise ( Moon set), calculated at the center of the visible disk of the Moon; *h* - height of the Moon, *δ* - declination of the Moon, *A* - the azimuth of the Moon.

| phenomenon | h, (0) | δ, (0) | A, (0) | The letter designation |
|---|---|---|---|---|
| northern major standstill moonrise | 0.35 | 28.99 | 44.94 | B1 |
| southern major standstill moonrise | 0.35 | -28.99 | 136.14 | B3 |
| northern minor standstill moonrise | 0.35 | 18.69 | 62.24 | C1 |
| southern minor standstill moonrise | 0.35 | -18.69 | 118.62 | C3 |
| northern major standstill moonset | 0.35 | 28.99 | 315.06 | - |
| southern major standstill moonset | 0.35 | -28.99 | 223.86 | - |
| northern minor standstill moonset | 0.35 | 18.69 | 297.76 | - |
| southern minor standstill moonset | 0.35 | -18.69 | 241.38 | - |
| equinox major standstill moonrise | 0.35 | 5.15 | 82.78 | B2 |
| equinox minor standstill moonrise | 0.35 | -5.15 | 97.99 | C2 |
| equinox major standstill moonset | 0.35 | 5.15 | 277.22 | - |
| equinox minor standstill moonset | 0.35 | -5.15 | 262.01 | - |

For fixation direction always requires two gnomons (viewfinders). We assumed that the role of one of the viewfinder could perform gnomon of analemmatic sundial installed in the central the well corresponding placement of the gnomon at the equinoxes (point O). The second viewfinder could be subject similar to the gnomon, that is installed directly into the wells located east analemmatic sundial. Lines corresponding to the calculated astronomically important directions (Table. 4, Tab. 5), were shown emanating from a central point O, corresponding to the first gnomon, in Figure 9.

---

[6] *The letter designation in Figure 9.*

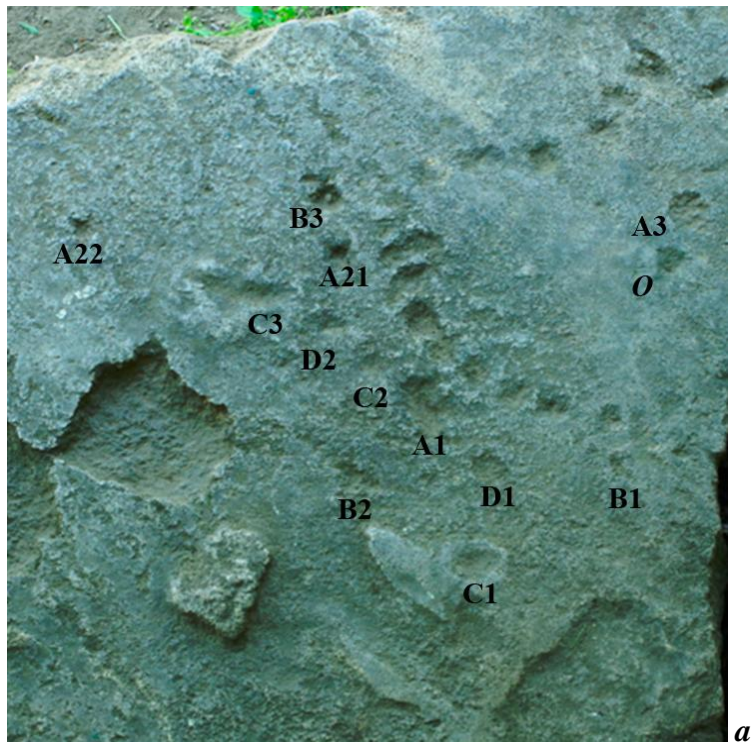

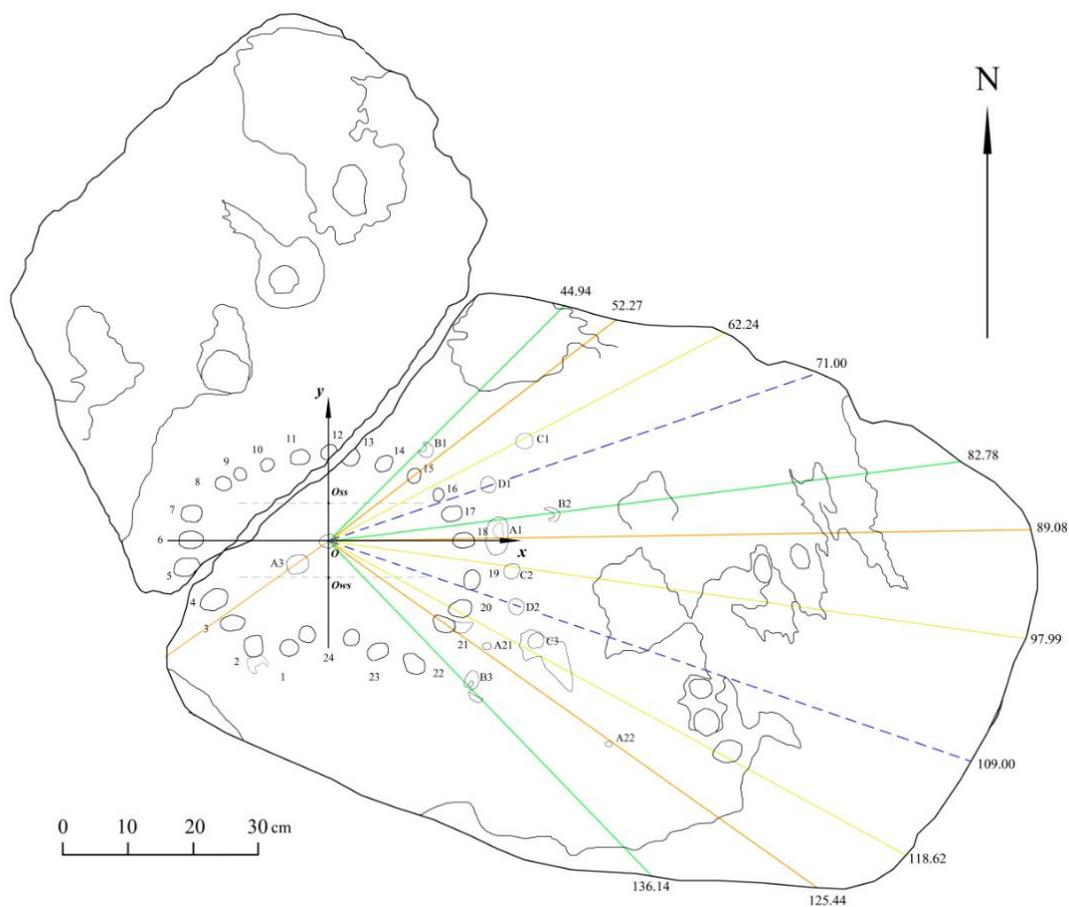

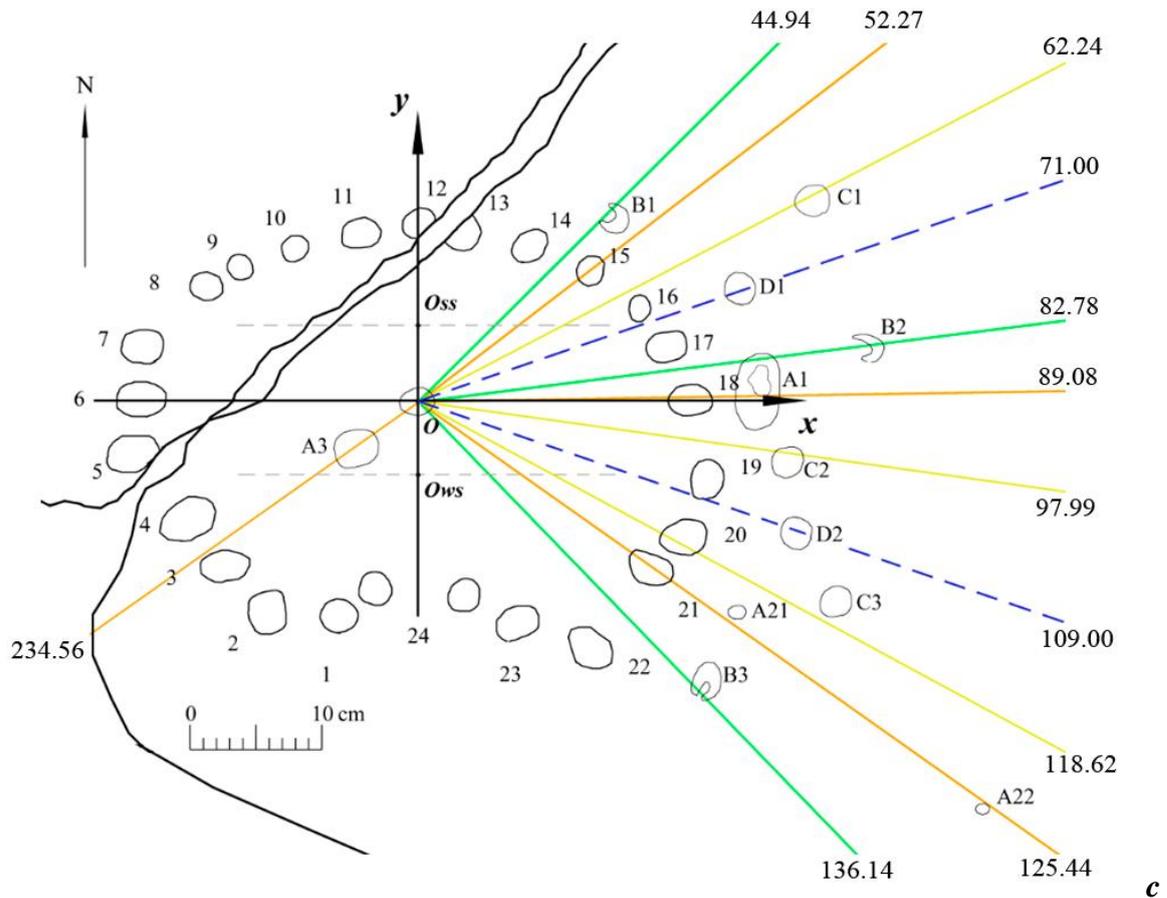

**Figure 9**. Kurgan field Tavriya - 1, kurgan 1, burial 2, slab with the wells: ***a*** – photo of fragment of slab with wells (view from NE), ***b*** – drawing of slab with surface defects and marked astronomically important directions, ***c*** – drawing of fragment of slab with wells and astronomically important directions. *N* - True North.

In Figure 9 index *A1* designated well corresponding to the azimuth of sunrise at the equinox, *A21* - sunrise at the winter solstice, *A22* - sunrise at the winter solstice, *A3* - sunset at the winter solstice (the sunrise at the summer solstice), *O* - the place of installation of analemmatic sundial gnomon at the equinox (one of the viewfinder), *B1* - the northernmost rise of High Moon, *B2* - rise of High Moon at the equinox, *B3* - the southernmost rise of High Moon, *C1* - the northernmost rise of Low Moon, *C2* – rise of Low Moon at the equinox, *C3* - the southernmost rise of Low Moon, *D1*, and *D2* - wells that have no correlation with the rise / set of luminaries. Astronomically important directions related to the Sun are marked in orange color, associated with the High Moon – green color, associated with the Low Moon – yellow color, with no connection to the rising luminaries - blue color.

Almost all of the lines, corresponding to the calculated us astronomically important directions and associated with the eastern sector of the sky, passing through wells located east of wells arranged in a circle. An exception is the line corresponding to the direction associated with the rising of the Sun near the summer solstice. However, approximately on continuing this line, but in the western sector of analemmatic sundial located well *A3*, which we compared with the direction of the sunset at the winter solstice. The line passing through the point *Ows*, where it is necessary to establish the gnomon of analemmatic sundial in the winter solstice, touches the edge of the well *A3*. This fact is further evidence of our interpretation of wells arranged in a circle on the slab as analemmatic sundial and auxiliary wells as astronomical markers.

By setting the second viewfinder to well *A3*, with a small margin of error, it is possible to observe objects in both directions: a sunset on the winter solstice (*OA3*) and sunrise at the summer solstice (*A3O*) (Fig. 9).

Wells A21 and A22 is less than the rest of the wells and their diameter does not exceed 1 cm. Line corresponding to the sunrise at the winter solstice, and passing through the center of analemmatic sundial, passes between the wells (see Fig. 9 c). At the expense of the shadows cast by gnomons, when installed in these wells at the time of sunrise on the slab will appear thin lighted strip passing through the center sundial. We believe that this feature could be used to select and original marking of the center sundial.

Because daylength in winter is the smallest and the height of the Sun in the upper culmination is minimal, the small size of these wells could symbolize the "little" Sun - the Sun - "grain", which began to grow (increase the height of the Sun in the upper culmination) and increase (increase light day) from the winter solstice. It is interesting that in the countryside among the Slavic peoples still widespread tradition of "sowing" - sprinkle by grain or oats on New Year or Christmas Day (December 25), which in time almost coincides with the winter solstice (December 21-22) .

Large well *A3*, correlated with the rising of the sun in the summer solstice (*A3O*), due to its location in the western sector (associated with the set of luminaries) may also symbolize the maximum height of the Sun in the upper culmination, and the maximum length of daylight that begin to decrease with the summer solstice, i.e. symbolize the beginning of "dying" of the Sun.

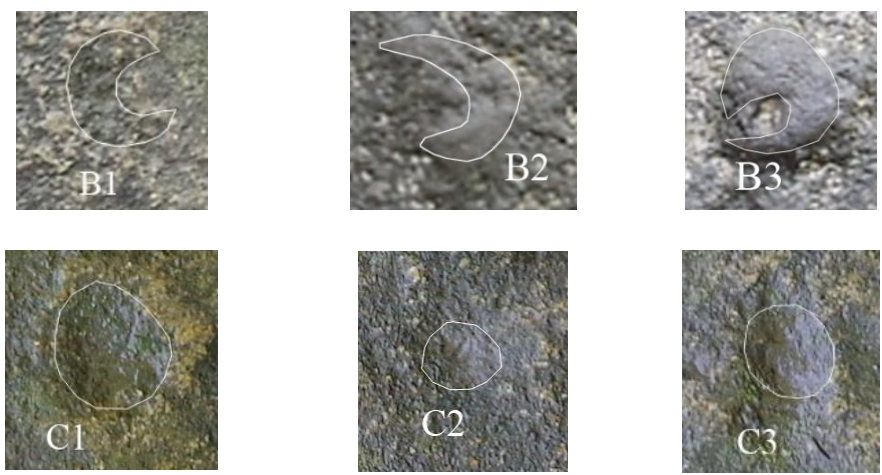

**Figure 10**. Kurgan field Tavriya - 1, kurgan 1, burial 2, slab with the wells, wells associated with the direction of the rise High (*B1, B2, B3*) and Low (*C1, C2, C3*) Moon. The white lines are rough contours of the wells.

Well *A1* associated with the sunrise at the equinox reminds short groove. Most likely, this is due to the rapid apparent motion of the Sun near the date of the equinox and, as a consequence, the difficulty in determining the time in practice (days) equinox and, accordingly, the exact azimuth of sunrise on the day of the equinox.

Wells crossed the lines associated with the High Moon, remind, from our point of view, in the form of a half moon. Wells associated with Low Moon are circular (see Fig. 10). It is unclear why a crescent shape compared with High Moon, but such a division in the form wells can be seen as further evidence of the existence of concepts of High and Low Moon among Srubna population.

During the study, we found that two wells D1 and D2 are not associated with any of the calculated astronomical directions. However, together with the adjacent wells, correlated with the astronomically important directions, they are arranged in a clearly visible line that, from our point of view, the evidence and their astronomical appointment.

Except the rising and setting luminaries, the brightest and most easily noticeable astronomical phenomenon is the change in the height of the Sun in the upper culmination of during a year. To test the hypothesis about the possible connection wells with this phenomenon, we measure an azimuth wells *D1* and *D2*: $A_{D1} \approx 71^0$, $A_{D2} \approx 109^0$ and analyzed the height of the Sun at the top of culmination in the days when the azimuth of sunrise coincides with the azimuths of these wells (Table. 6). For dates with a given azimuth of sunrise were calculated heights of the Sun at the top of the culmination. The calculations were performed for the 1200 BC and 2014 AD using astronomical program RedShift-7 Advanced for geographic coordinates of Tavriya-1.

**Table 6.** Height of the Sun at the top of the culmination. *A* - the azimuth of sunrise, *h* - height of the Sun at the top of the culmination.

|   | Ai, (0) | h, (0) | astronomical event | date for 1200 BC | date for 2014 AD |
|---|---------|--------|--------------------|------------------|------------------|
| 1 | 125.44 | 18.9 | winter solstice | 31.12 | 22.12 |
| 2 | 109.00 | 29.7 | - | 25.02 | 14.02 |
| 3 | 89.08 | 42.8 | vernal equinox | 01.04 | 20.03 |
| 4 | 71.00 | 55.0 | - | 04.05 | 23.04 |
| 5 | 52.27 | 66.5 | summer solstice | 04.07 | 21.06 |
| 6 | 71.00 | 55.0 | - | 02.09 | 21.08 |
| 7 | 89.08 | 43.0 | autumnal equinox | 04.10 | 23.09 |
| 8 | 109.00 | 29.3 | - | 07.11 | 28.10 |

Analyzing the calculated height of the sun in the upper culmination, we identify patterns associated with the wells D1 and D2. Height of the Sun at the top of culmination, in the days when the Sun rises in the direction with an azimuth of wells D1 and D2, is equal to the average value between the heights of the nearest preceding and succeeding the solstice and equinox with an error $\approx 1^0$.

For example, 02/25/1200 BC (14/02/2014 AD) Sun in the upper culmination has a height $h_{D2}$=29.7$^0$, height of the Sun in the upper culmination of the winter solstice $h_{ws}$=18.9$^0$, and the vernal equinox $h_{eq}$=42.8$^0$. Then $(h_{ws}+h_{eq})/2$=30.9$^0$≈$h_{D2}$. And at 04/05/1200 BC (23/04/2014 AD) Sun in the upper culmination has a height $h_{D1}$=55$^0$. Height of the Sun in the upper culmination of the summer solstice $h_{ss}$=66.5$^0$, and at the spring equinox $h_{eq}$=42.8$^0$. Then $(h_{eq}+h_{ss})/2$=54.7$^0$≈ $h_{D1}$.

Accommodation in immediate proximity to analemmatic sundial associated with it the group of markers of sunrise luminaries could be used to determine the direction of movement of the gnomon of analemmatic sundial for it proper functioning during the year. For example, the time of the summer solstice could be determined using wells *A3* and *O*. In this case, if at 6 am at the equinox (the azimuth of the Sun A = 89.08$^0$) the shadow of the gnomon, installed at the point *O*, fell to the well of the 6 hours, then in the summer solstice at 6 am (azimuth of the Sun A = 73.27$^0$) the shadow of the gnomon has fallen between wells 4 and 5 hours (see Fig. 11). In order to sundial shows the correct time, it was necessary to move the gnomon to the point *Oss*. Similar movements of the gnomon were necessary to carried on throughout the year.

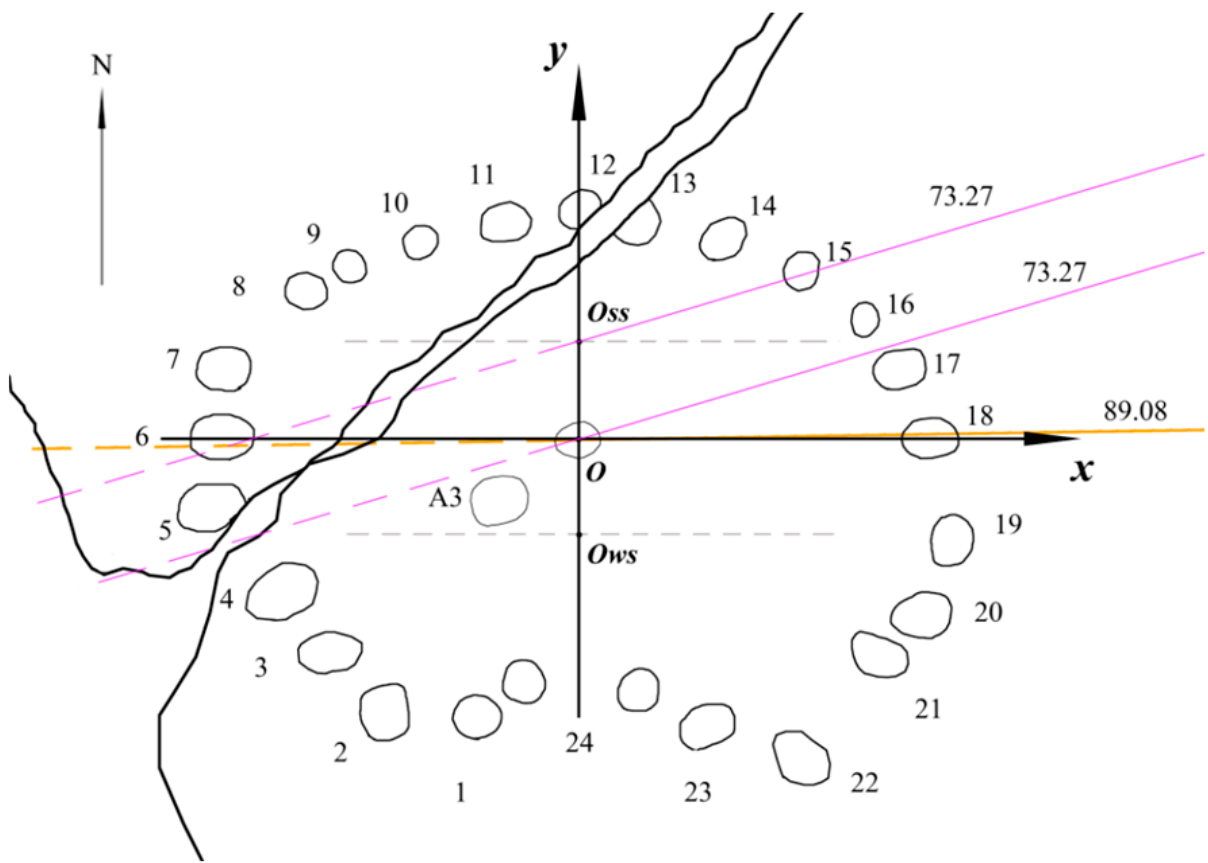

**Figure 11**. Kurgan field Tavriya - 1, kurgan 1, burial 2, drawing of fragment of slabs with wells. The solid lines indicate the direction to the azimuth of the Sun at 6 am at the day of the equinox (orange line) and the summer solstice (pink line). Dotted lines indicate the shade of gnomons the same time. N - True North.

Simultaneous placement group of wells, associated with a sundial, and a group of wells - markers of astronomically significant directions on one plate can be considered as an interim stage of development analemmatic sundial and, therefore, as possible evidence of local origin and the very idea of analemmatic sundial, and measurement technology time with it.

Detection on the slab including astronomical markers wells associated with apparent motion of the Moon, may also be evidence of an attempt to use wells arranged in a circle, as the moondial. However, the simplest moondial, based on analemmatic sundial, can show the correct time on the night of the full moon, while moving the gnomon to the point *Ows* near the summer solstice, to the point *Oss* near the winter solstice, and to the point *O* - near the equinox, only. In this case the wells in the work part of the "dial" will fit to not by day, but night hours. Well of 6 hours corresponds to 18 hours, well 12 hours - 24 hours, as well of 18 hours - 6 hours. All lunar direction marked wells - astronomical markers at the slab, connected with the rising of the Moon during the full moon, which can be considered as evidence for the assumption that the use of a slab from Tavriya-1 as moondial.

The wells of the second circle, clearly visible on the image of slab in 1991 (see Fig. 1) were not interpreted. This is due to the fact that modern extent of damage of the appropriate the slab edge are not allowed to make them reliable analysis. Could not we compare with astronomical directions and three large bowl-shaped pits on the opposite edge of the slab. Despite their good state of preservation, the line on which they are located, does not coincide with sufficient accuracy with calculated astronomical directions. About it can be compared with the direction to the sunrise at the summer solstice and sunset at the winter solstice. In many traditions these periods were considered as holidays, so perhaps three large wells functioned as containers for festive sacrifices.

**Conclusion**

Thus, as a result of our study was analyzed the spatial arrangement of wells on the slab of Srubna burial of kurgan field Tavriya-1. During research identified two interrelated groups of wells. The first group of wells was interpreted by us as analemmatic sundial, which could work as a moondial at the night of the full moon, and the second group - as markers of astronomically significant directions, support and for sundial, and for moondial. We have developed and described the way in which in ancient times could determine the moments of solstices and equinoxes using analemmatic sundial gnomon and mobile gnomons installed in wells belonging to the second group. The slab with wells from Tavriya-1 allows following the movement of the Moon and the Sun, to determine the time of the solstices and equinoxes, to measure the time during the day using analemmatic sundial and at night under a full moon with the help of moondial. Communication wells on the slab from Tavriya-1 with astronomically important directions and the definition of time by analemmatic sundial and moondial, allows us to determine the slab with wells from Tavriya-1 as the oldest astronomical instrument, the main part of which is a sundial.

**Acknowledgements**

The authors are sincerely thank to Vera Alekseevna Larenok and Valery Fedorovich Chesnok for the support of research, and we wish to thank to administration and employees of the Archaeological Museum "Tanais" for storing many years of unique slab from kurgan field Tavriya-1.